 \documentclass[12pt,onecolumn]{IEEEtran}
 \usepackage[colorlinks,bookmarksopen,bookmarksnumbered,citecolor=red,urlcolor=red,]{hyperref}
 \usepackage{indentfirst}
 \usepackage{algorithm,algorithmic,amsbsy,amsmath,amssymb,epsfig,bbm,mathrsfs, bbm} 
 \usepackage{multirow}
 \usepackage{mathrsfs}
 \usepackage{epstopdf}
 \usepackage[subfigure]{graphfig}
 \usepackage{bm}
 \usepackage{cite}
 \usepackage{verbatim}
 \usepackage{amsfonts}
 \usepackage{amsthm}
 \usepackage{geometry}
\begin{document}
\title{ 
Ambient Backscatter Networking: A Novel Paradigm to Assist Wireless Powered Communications
}

\author{ 
Xiao Lu, Dusit Niyato, Hai Jiang, Dong In Kim, Yong Xiao and Zhu Han
\thanks{Dong In Kim is the corresponding author} 
 }
\maketitle

\begin{abstract}
Ambient backscatter communication technology has been introduced recently, and is then quickly becoming a promising choice for self-sustainable communication systems as an external power supply or a dedicated carrier emitter is not required. By leveraging existing RF signal resources, ambient backscatter technology can support sustainable and independent communications and consequently open up a whole new set of applications that facilitate Internet-of-Things (IoT). In this article, we study an integration of ambient backscatter with wireless powered communication networks (WPCNs). We first present an overview of backscatter communication systems with an emphasis on the emerging ambient backscatter technology. Then we propose a novel hybrid transmitter design by combining the advantages  
of both ambient backscatter and wireless powered communications. Furthermore, in the cognitive radio environment, we introduce a multiple access scheme to coordinate the hybrid data transmissions. The performance evaluation shows that the hybrid transmitter outperforms traditional designs. In addition, we discuss some open issues related to the ambient backscatter networking.
\end{abstract}
 
\section{Introduction}

Information transmission based on modulated backscatter of incident signals from external RF sources has emerged as a promising solution for low-power wireless communications. The power consumption of a typical backscatter transmitter is less than 1 $\mu $W \cite{B.2014Kellogg}, which renders excessively long lifetime, e.g., 10 years, for an on-chip battery. This low power consumption well matches the harvestable wireless energy from RF sources, e.g., typically from 1 $\mu$W to tens of $\mu$W~\cite{XLuSurvey,X.2015Lu}. This additionally renders RF energy harvesting to be an alternative to power backscatter transmitters. Furthermore, backscatter communications can be embedded into small gadgets and objects, e.g., a radio frequency identification (RFID) and passive sensor. Therefore, backscatter communications is also envisioned as the last hop in the Internet-of-Things (IoT)~\cite{V.2013Liu}, which requires low cost and ubiquitous deployment of small-sized devices~\cite{D.2012Miorandi}.
 

Due to recent dramatic increases in application demands, the requirement for backscatter communications has gone beyond the conventional RFID towards a more data-intensive way.  This strongly raises the need for re-engineering backscatter transmitters for better reliability, higher data rates, and longer interrogation/transmission range. However, traditional backscatter communication techniques, e.g., RFID, are hindered by three major shortcomings: 1) The activation of backscatter transmitters relies on an external power supply such as an active interrogator (also called a reader or carrier emitter) which is costly and bulky. 2) A backscatter transmitter passively responds only when inquired by a reader. The communication link is restricted in one hop, typically with the distance ranging from a few centimeters to a few meters. 3) A backscatter transmitter's reflected signal could be severely impaired by adjacent active readers, significantly limiting the device usage in a dense deployment scenario.

Recently, ambient backscatter communication technology~\cite{V.2013Liu} has emerged to overcome some of the above limitations. An ambient backscatter transmitter functions using ambient RF signals for both backscattering and batteryless operations through RF energy harvesting. 
Energy is harvested from ambient RF sources in the radio environment, avoiding the development and deployment cost of readers 
as well as improving the flexibility in use. Despite many benefits, the study of ambient backscatter communications is still at its nascent stage. Various challenges arise from the data communication and networking perspectives~\cite{X.2017LuVTC}. This motivates our work in this article. 


\section{Backscatter Communications}

This section first describes the backscatter communications from the system perspective and then reviews the fundamental principles of backscatter communications. 

\subsection{Backscatter Communication Systems}

Following the standard communication terminology, a backscatter communication system has three major components, i.e., a carrier emitter, a backscatter transmitter, and a receiver. The carrier emitter can be either a dedicated RF generator that only broadcasts continuous sinusoidal waves~\cite{J.2014Kimionis} or an ambient RF transmitter communicating with its own intended receiver(s)~\cite{V.2013Liu}, e.g., a Wi-Fi router~\cite{B.2014Kellogg}. A backscatter transmitter is a device that backscatters a signal from the carrier emitter for data transmission to the backscatter receiver. The receiver is capable of decoding the data from the modulated backscatter signal. In an RFID context, the carrier emitter and receiver can be co-located, and it is called an interrogator or reader.

Backscatter communication systems can be classified based on the type of power supply. 
\begin{itemize}
	\item In passive systems, the backscatter transmitter relies on the exogenous RF waves from a carrier emitter for both operational power and a carrier signal to reflect. To communicate, the backscatter transmitter first harvests energy from incident RF waves through rectifying, typically by a rectenna or charge pump. Once the rectified DC voltage exceeds a minimum operating level, the device is activated. 
	The device then tunes its antenna impedance to generate modulated backscatter using the instantaneously harvested energy. Therefore, passive systems are featured with coupled and concurrent backscattering and energy harvesting processes.
	For example, the experiment in~\cite{D.2016Assimonis} demonstrates that a batteryless backscatter sensor can function continuously with an input RF power of 18 dBm (equivalently, 0.1103 $\mu$W/cm$^{2}$ power density) for energy harvesting.	
	Because part of incident RF waves are utilized for energy harvesting to attain the operational power, the effective transmission range of a passive device is relatively short, typically within a few meters. Moreover, it has limited data sensing, processing, and storage capabilities. Nevertheless, the key benefit of the passive device is its low cost and small size. For example, the recent advance in chipless implementation such as surface acoustic wave (SAW) tag~\cite{V.2010Plessky} allows fabricating a passive device with the cost of only 10 cents.  
	\item In semi-passive systems, the backscatter transmitter, instead of relying on the harvested energy from the carrier emitter alone, is equipped with an internal power source. 
	Thereby, without needing to wait until harvesting enough energy, the access delay is significantly reduced. 
	The semi-passive device enjoys better reliability and wider range of accessibility, typically up to tens of meters. Another benefit is that the battery can support larger data memory size and on-chip sensors~\cite{J.2010Yin}, which remarkably widens the functionality and usability of the device. However, the semi-passive device still utilizes the RF signals from a carrier emitter for backscattering, and thus, does not largely improve the transmission rate over its passive counterpart.
	Moreover, the semi-passive device has limited operation time subject to the battery capacity.
\end{itemize}

\begin{figure}
\centering
\includegraphics[width=0.8\textwidth]{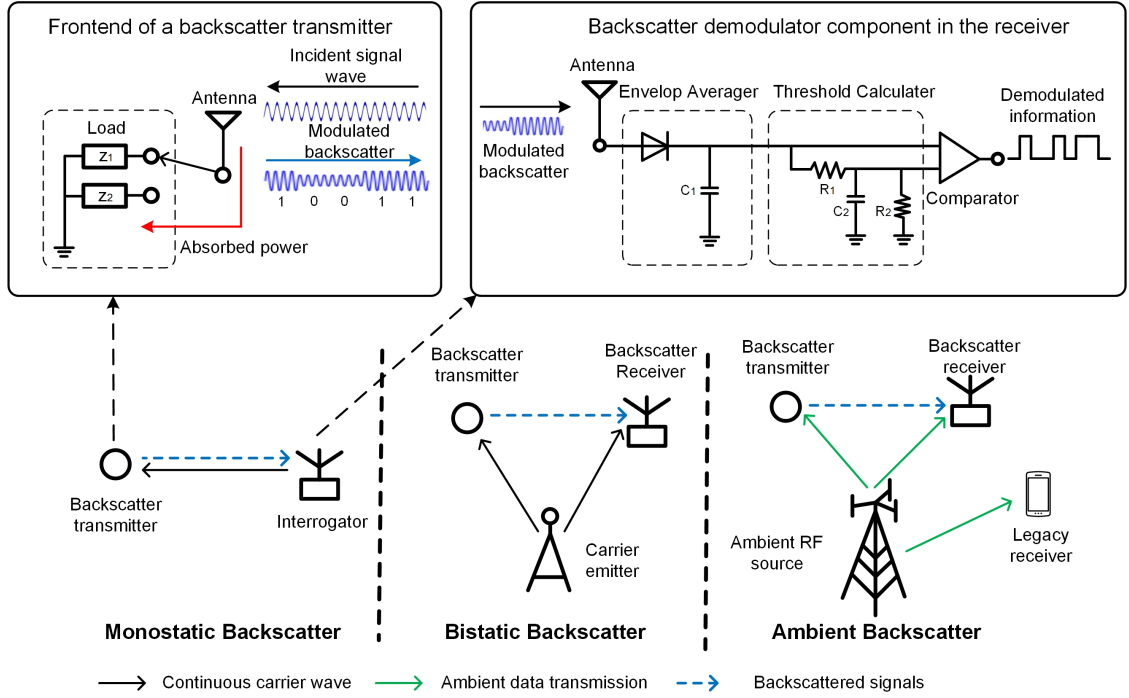} 
\caption{Paradigms for backscatter communications.} \label{fig:Backscatter_systems} 
\end{figure} 
 
Figure~\ref{fig:Backscatter_systems} illustrates three configurations of backscatter communication systems.
\begin{itemize}
	\item {\em Monostatic backscatter configuration} consists of two components, i.e., an interrogator and a backscatter transmitter, e.g., an RFID tag. The interrogator as the carrier emitter first releases RF signals to activate the backscatter transmitter. Once activated, the backscatter transmitter performs modulation utilizing the same RF signals from the interrogator. The reflected modulated backscatter signals are then captured by the interrogator which acts as the backscatter receiver. Since the carrier emitter and the backscatter receiver are co-located, the backscattered signal suffers from a round-trip path loss~\cite{J.2014Kimionis}. 
	The monostatic configuration is mostly adopted for short-range RFID applications. 

	\item {\em Bistatic backscatter configuration} differs from the monostatic counterpart in that the interrogator is replaced by a separate carrier emitter and a separate backscatter receiver. 
	The bistatic configuration allows setting up more flexible network topologies. For example, a carrier emitter can be placed at an optimal location for backscatter transmitters and receivers. Moreover, the bistatic configuration can mitigate the doubly near-far effect~\cite{X.Lu2016} by distributing several carrier emitters in the field. It was shown that the fabrication cost of the carrier emitter and backscatter receiver in the bistatic configuration is cheaper than that of the interrogator in monostatic configuration due to less complex design~\cite{J.2014Kimionis}.

	\item {\em Ambient backscatter configuration} is similar to the bistatic configuration. However, the carrier emitter is an existing RF source, e.g., a TV tower, base station, or Wi-Fi access point, instead of a dedicated device. Unlike the bistatic backscatter configuration in which the communication is always initiated by the carrier emitter, the ambient backscatter transmitter can harvest RF energy from the ambient RF source and initiate transmission to its receiver autonomously. 
	Nevertheless, compared with the bistatic configuration, the transmission range of an ambient backscatter transmitter is limited because the performance is affected by channel conditions and by natural variations of the ambient signals~\cite{D.2015Bharadia}. 

\end{itemize} 

\begin{table} \footnotesize 
\centering
\caption{Comparison of Ambient Backscatter Communication Prototypes.} \label{prototypes}
\begin{tabular}{|p{2cm}| p{2cm}|p{3cm}|p{3.2cm}|p{2.2cm}|p{1.5cm}|p{1.2cm}|}
\hline
\footnotesize {\bf Reference} & {\bf Energy Source } & {\bf Rate $@$ Operating Range } & {\bf Energy Consumption/ Efficiency } & {\bf Center Operating Frequency} & {\bf Bandwidth} & {\bf Year} \\ 
\hline
\hline
Ambient backscatter prototype~\cite{V.2013Liu} & Ambient TV signals & Outdoor: 1 kbps $@$ 2.5 feet; Indoor: 1 kbps $@$ 1.5 feet & Transmitter: 0.25 $\mu$W; Receiver: 0.54 $\mu$W & 539 MHz & 50 MHz & 2013 \\
 \hline
 $\mu$mo prototype \cite{N.2014Parks} & Ambient TV signals and/or solar power & 1 Mbps @ 7 feet & Receiver: 422 $\mu$W & 539 MHz & 1 MHz & 2014 \\
 \hline 
 $\mu$code \cite{N.2014Parks} & Ambient TV signals and/or solar power & 333 bps @ 30 feet& Receiver: 8.9 $\mu$W & 539 MHz & 1 MHz & 2014 \\
 \hline 
Wi-Fi backscatter prototype \cite{B.2014Kellogg} & Ambient Wi-Fi signals & Downlink: 20 kbps @ 2.2 meters; 5 kbps @ 3 meters; & Transmitter: 0.69 $\mu$W; Receiver: 9.0 $\mu$W & 2.4 GHz & 20 MHz & 2014 \\
\hline
BackFi prototype \cite{D.2015Bharadia} & Ambient Wi-Fi signals & 5 Mbps $@$ 1m;
 1 Mbps $@$ 5 m & 3.15 petaJoule/bit & 2.4 GHz & 20 MHz & 2015 \\
\hline
\end{tabular}
\end{table} 

In addition to lower cost and less energy consumption, the ambient backscatter configuration does not need extra spectrum to operate. In the other configurations, the use of interrogators or dedicated carrier emitters costs not only additional power but also frequency
resources.
Besides, both the emitted signal from interrogators (or dedicated carrier emitters) and the reflected signal from backscatter transmitters can cause severe interference to other wireless devices, especially in a dense deployment scenario, e.g., large supply chains, and when high-gain antennas are employed. 
By contrast, 
the ambient backscatter signal does not cause any noticeable interference to other wireless devices unless their distance to the ambient backscatter transmitter is extremely close, e.g., less than 7 inches
as demonstrated in a real experiment in \cite{V.2013Liu}.
Thus, the choice of working frequency of an ambient backscatter transmitter can be determined based on the specific applications. For example, the outdoor implementation can be based on TV frequency~\cite{V.2013Liu} while Wi-Fi frequency \cite{D.2015Bharadia,N.2014Parks} would be an appropriate option for indoor applications.
Interestingly, a recent study in~\cite{D.1605.04805Darsena} reveals that ambient backscatter communications can even improve the legacy receiver's performance if the legacy receiver can leverage the additional diversity from the backscattered signals.
Moreover, in monostatic and bistatic backscatter configurations, all the backscatter transmitters need to be located within the coverage of their dedicated interrogators or carrier emitters in order to receive RF signals and to be scheduled, and thus, usually only one-hop communication is feasible. Differently, multiple ambient backscatter transmitters can initiate transmissions independently and simultaneously, thus making multi-hop communications possible to overcome the short-range issue of backscatter links.
Table~\ref{prototypes} summarizes and compares the important ambient backscatter communication prototypes. 



\subsection{Fundamentals}
\subsubsection{Basic principles of modulated backscatter}

Different from other conventional wireless communication systems, a backscatter transmitter does not generate active RF signals. Instead, the backscatter transmitter maps a sequence of digital symbols onto the RF backscattered waveforms at the antenna. The waveform adaptation is done by adjusting the load impedance, i.e., reflection coefficient, of the antenna to generate different waveforms from that of the original signal. This is known as load modulation. Figure~\ref{fig:Backscatter_systems} shows the diagram of a backscatter transmitter with binary load modulation. 
It has two loads with the impedances intentionally matching and mismatching with the antenna impedance, respectively. 
The antenna reflection coefficient, and thus the amount of the reflected signal from the antenna\footnote{The reflected signal is composed of two parts: structural mode and antenna mode scatterings~\cite{D.2010Dardari}. The former is determined by the antenna physical property, e.g., material and shape. The latter is a function of antenna load impedance.}, can be tuned by switching between the two impedance loads.
Specifically, 
when the load with the matched impedance is chosen, most of the incident signal is harvested, i.e., an absorbing state. Conversely, if the antenna switches to the other load, a large amount of the signal is reflected, i.e., a reflecting state.   
A backscatter transmitter can utilize an absorbing state 
and a reflecting state to indicate a bit ``0" and a bit ``1'', respectively, to its intended receiver.
It follows that, in the reflecting state, the receiver will observe a superposition of the original wave from the signal source (e.g., the interrogator) and the backscatter transmitter's reflected wave. In the absorbing state, the receiver will only see the original wave. The states are then interpreted as information bits.



The information rate can be adapted by varying the bit duration. As the backscatter transmitter only works as a passive transponder that reflects part of the incident signals during modulation, the implementation can be very simple and requires no conventional communication components such as oscillator, amplifier, filter, and mixer. Typically, the backscatter transmitter only consists of a digital logic integrated circuit, an antenna, and an optional power storage, making it cheap, small, and easy to deploy.
 
 
\subsubsection{Modulation and demodulation}

In backscatter communications, information is usually carried in amplitude and/or phase of the reflected signals. 
Binary modulation schemes such as amplitude shift keying (ASK) and phase shift keying (PSK) are most commonly adopted in passive backscatter systems. However, due to low spectrum efficiency, binary modulations result in a low data rate, e.g., up to 640 kbps for FM0 and Miller coding. This calls for research efforts in multilevel modulation, i.e., higher-order constellations, to accommodate demand from data-intensive applications. Recent studies have shown that backscatter systems can support $M$-ary quadrature amplitude modulation (QAM), e.g., 16-QAM~\cite{J.2012Thomas} and 32-QAM~\cite{A.2015Shirane} as well as $N$-PSK, e.g., 16-PSK~\cite{D.2015Bharadia}. A passive RF-powered backscatter transmitter~\cite{A.2015Shirane} operating on 5.8 GHz can achieve 2.5 Mbps data rate with a 32-QAM modulation at a ten-centimeter distance. Moreover, 4-PSK and 16-PSK modulations have been implemented in an ambient backscatter prototype, i.e., BackFi~\cite{D.2015Bharadia}. The experiments demonstrate 5 Mbps data rate at the range of one meter. 



Because of amplitude and/or phase modulation, a backscatter receiver needs to determine the amplitude and/or phase change. It is relatively simpler to implement an amplitude demodulator. Figure~\ref{fig:Backscatter_systems} again shows the diagram of binary amplitude demodulation based on envelope detection at the receiver. The circuit has four components: an antenna, an envelope averager, a threshold calculator, and a comparator. 
The instantaneously incoming waves at the antenna are smoothed at the envelope averager, which gives an envelope of the instantaneous signals. The threshold calculator generates a threshold value by taking the mean of the long-term averaged envelope.
Then, the comparator makes a comparison between the smoothed instantaneous envelope of the received modulated backscatter and the threshold to decide the value of the information bits. 
Alternatively, demodulating from backscattered waves with phase variation requires phase detection. 
Some most common practices of phase demodulation include the adoption of a homodyne receiver with an RF in-phase/quadrature demodulation~\cite{J.2012Thomas} and channel estimation~\cite{D.2015Bharadia}. For example, BackFi~\cite{D.2015Bharadia} uses a preamble to estimate the combined forward-backward channel. This estimated value of the channel is then used for decoding the information bit modulated on the phase.

\section{Ambient Backscatter Assisted Wireless Powered Communication Networks} \label{Sec:BA-WPC}

Wireless powered communication is an active radio transmission paradigm powered by RF energy harvesting~\cite{XLuSurvey}. 
The recent development in the circuit technology has greatly advanced an RF energy harvester implementation in terms of hardware sensitivity, RF-to-DC conversion efficiency, circuit power consumption, etc. This has made powering active radio communications by ambient RF energy harvesting~\cite{I.May2015Flint,X.March2015Lu,I.December2014Flint} become feasible. For example, a recent design and implementation of RF-powered transmitter~\cite{J.2015Kim}  showed that the transmission rate could reach up to 5 Mbps.

We propose a novel hybrid transmitter that combines ambient backscatter communications and the wireless powered communications, both of which can harvest RF energy and support self-sustainable communications~\cite{V.2013Liu,X.May2016Lu}.  On the one hand, 
a wireless powered transmitter requires a long time to accumulate enough energy for active transmission when the ambient RF signals are not abundant~\cite{D.2015Niyato}. In this circumstance, ambient backscattering may still be employed as it has ultra-low power consumption.
On the other hand, since the ambient RF signals may alternate between ON state (i.e., the signals are present) and OFF state (i.e., the signals are absent), ambient backscattering cannot be adopted when the RF signals are OFF. In this case, a wireless powered transmitter can be adopted to perform active RF transmission using energy that was harvested and stored before when the RF signals were ON~\cite{X.June2014Lu,D.May2016Niyato,D.June2015Niyato}.
Therefore, these two technologies can well complement each other and result in better data transmission performance.

\begin{figure}
\centering
\includegraphics[width=0.65\textwidth]{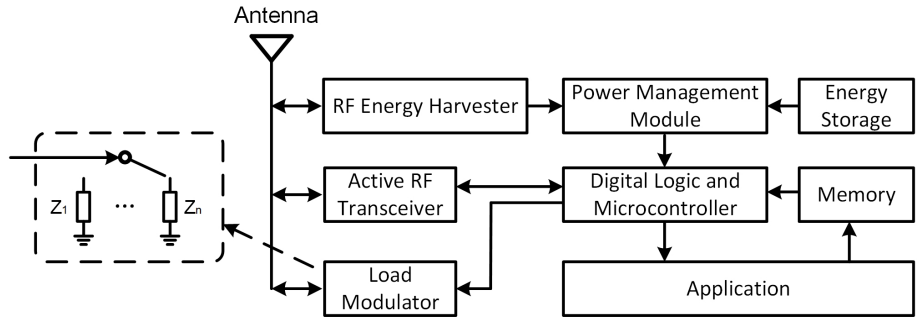} 
\caption{The structure of the hybrid transmitter} \label{fig:integrated_circuit} 
\end{figure}

Figure~\ref{fig:integrated_circuit} illustrates the block diagram of the hybrid transmitter. The transmitter consists of the following components: an antenna, a power management module, an energy storage, a load modulator, an RF energy harvester, an active RF transceiver, a digital logic and microcontroller, and a memory storage connected to an external application (e.g., a sensor). 
We can see that many circuit components such as the RF energy harvester, memory and energy storage, can be shared between ambient backscatter and wireless powered communications.
Based on the state-of-the-art circuit technology for ambient backscatter transmitters~\cite{D.2015Bharadia,N.2014Parks} and wireless-powered transmitters~\cite{X.Lu2016,X.April2015Lu}, the proposed design for the hybrid transmitter may not require a considerably more complex hardware implementation than that for performing ambient backscatter or wireless powered transmission alone. 

The proposed design here 
allows a highly flexible operation to perform RF energy harvesting, active data transmission/reception, and backscattering. Such integration in the proposed design has the following advantages.
\begin{itemize}
	\item It supports a long duty cycle. When the hybrid transmitter has data to transmit, but not enough energy to perform active RF transmission, the device can perform backscattering for urgent data delivery. Since ambient backscatter communications can use instantaneously harvested energy, it does not consume the energy in the storage reserved for active RF transmission. Consequently, the duty cycle is improved significantly. Additionally, the delay to respond to the data transmission request is much shorter.  

	\item In comparison with an ambient backscatter transmitter, the hybrid transmitter can achieve a longer transmission range using active RF transmission when necessary.

	\item The hybrid transmitter is capable of offloading transmission from active RF broadcast to passive backscattering, thus alleviating interference issues. This is especially beneficial in a dense/ultra-dense network with high spatial frequency reuse as 1) there are various signal sources to facilitate backscattering and 2) ambient backscattering does not cause noticeable interference to many other users.

	\item The hybrid transmitter can still be used even without licensed spectrum, e.g., in cognitive radio networks. Coexisting with a primary user who is assigned a licensed channel, the hybrid transmitter can harvest energy and perform backscattering when the primary user is transmitting. When the primary user is idle, the hybrid transmitter can use active RF transmission to access the channel as a secondary user.

\end{itemize}

Despite the above advantages of the hybrid transmitter, some technical issues arise. With a single antenna setting, the hybrid transmitter cannot backscatter, harvest energy, and perform active data transmission simultaneously. Therefore, the transmitter has to determine when and how long each operation mode should be activated to achieve the optimal tradeoff among RF energy harvesting, ambient backscattering, and active RF transmission. The problem becomes more complicated with multiple transmitters in the network, which will be discussed next.

\section{Multiple Access Scheme for the Hybrid Transmitters in Cognitive Radio Networks} 

In this section, we aim to examine the proposed hybrid transmitter in a cognitive radio network scenario where there exists no licensed channel for the hybrid transmitter. To coordinate the hybrid data transmissions of multiple devices, we devise a multiple access scheme for the ambient backscatter assisted wireless powered communications. 
\subsection{System Model}

We consider a cognitive radio network consisting of a primary transmitter (PT), multiple secondary transmitters (STs), which are the hybrid transmitters as described in Section~\ref{Sec:BA-WPC}, and a common secondary access point (SAP), e.g., gateway. 
The SAP works as both the controller for coordinating data transmissions and the data sink for collecting data from the multiple STs, e.g., sensors. In order to decode the hybrid transmission from the STs, the SAP is equipped with both a conventional radio receiver and a backscatter receiver. When the PT transmits data on the licensed channel, the STs can backscatter for data transmission to the SAP or harvest energy to replenish its battery~\cite{V.2013Liu}. When the PT is not transmitting, i.e., the channel is idle, the STs can perform active data transmission to the SAP over the channel with the stored energy. 

\begin{figure}
\centering
\includegraphics[width=0.98\textwidth]{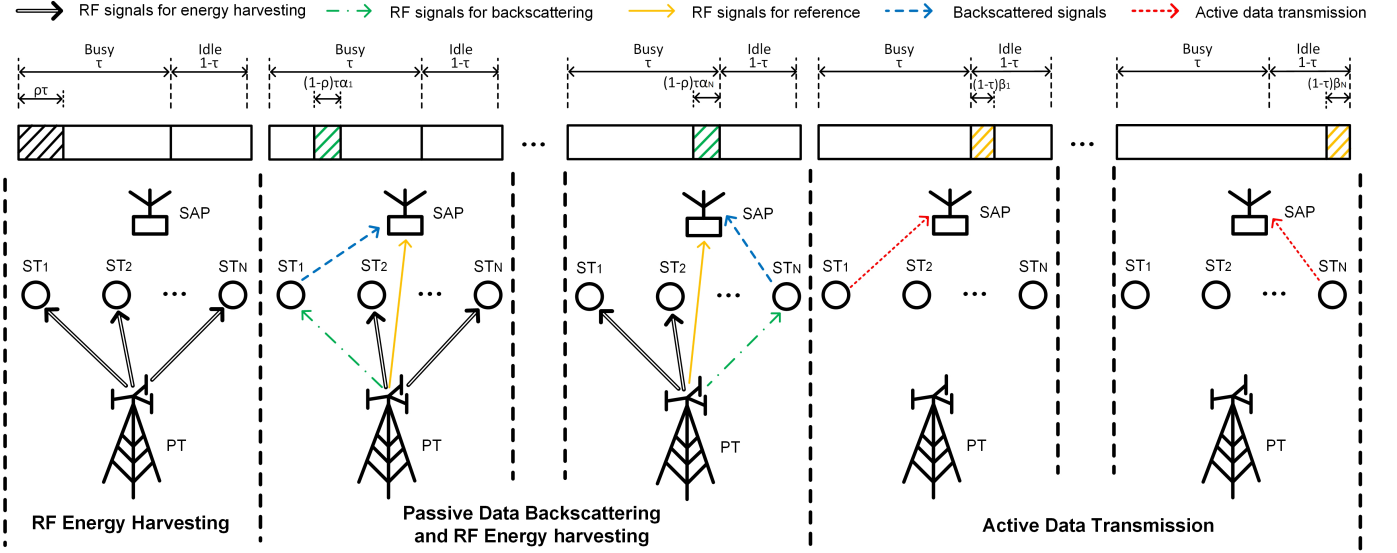} 
\caption{A multiple access scheme for ambient backscatter assisted wireless powered communication network.} \label{fig:MAC} 
\end{figure}

The network requires a multiple access scheme for the STs and SAP, which is proposed as shown in Figure~\ref{fig:MAC}. 
Let $\mathcal{N}=\{1, 2, \cdots, N\}$ denote the set of the STs. Let  $P_{T}$ and $\tau$ denote the transmit power and the normalized transmission period of the PT, respectively. The normalized channel idle period is then represented as $1-\tau$. The first part of the channel busy period is an energy harvesting subperiod with time fraction $\rho$. During this subperiod, all the STs harvest energy from the PT's signal to be used for later active data transmission. Then the time fraction $1-\rho$ during the channel busy period is used for backscattering. Let $\alpha_{n}$ denote the time fraction assigned to ST$_n$ for backscattering and $\sum_{n \in \mathcal{N}}\alpha_{n}=1$. When an ST is backscattering, the other STs can also harvest energy from the PT's signal. Thus, the total energy harvesting duration for ST$_n$ is  $[\rho+(1-\rho)\sum_{m \in \mathcal{N} \backslash  \{n\}} \alpha_{m}]\tau$. 
Let $h_{n}$ and $\mu$ denote the channel gain coefficient between the PT and ST$_n$, and RF-to-DC energy conversion efficiency of the STs, respectively.
The total amount of harvested energy by ST$_{n}$ can then be computed as $E^{H}_{n}=P_{T} h_{n}\mu \tau \left[ \rho + (1-\rho)\sum_{m \in \mathcal{N}\backslash\{n\}} \alpha_{m} \right]$.
During the channel idle period, ST$_n$ can perform active data transmission for a time fraction, denoted by $\beta_{n}$, in a sequential fashion, where $\sum_{n \in \mathcal{N}}\beta_{n}=1$. Evidently, there exists a tradeoff among energy harvesting, backscattering and active transmission. The throughput of the STs can be optimized by exploring this tradeoff. Let $\bm{\alpha}=\{\alpha_{1},\alpha_{2},\ldots, \alpha_{N}\}$ denote the set of the time fractions for backscattering and $\bm{\beta}=\{\beta_{1},\beta_{2},\ldots, \beta_{N}\}$ denote the set of time fractions for active data transmission of the STs.  The objective of our proposed multiple access scheme is to explore the optimal combination of $\rho$, $\bm{\alpha}$ and $\bm{\beta}$ to maximize the sum throughput of the STs.

\subsection{Multiple Access Control}

The transmission rate of backscattering depends on the physical configuration of the circuit~\cite{V.2013Liu}. Through the real implementation in \cite{V.2013Liu}, an ST can backscatter by load modulation without consuming energy from its battery. When the PT is transmitting and an ST is not scheduled for backscattering, it can harvest energy, part of which, denoted by $E_{C}$, is consumed by the circuit components. The surplus energy is stored in the battery and reserved for future active data transmission. We consider two important requirements in this multiple access scheme. One is the quality-of-service (QoS) requirement that the minimum throughput of each ST should be greater than or equal to the demand threshold, denoted by $R_{t}$. The other is the physical constraint that, during active data transmission, the allocated transmit power of ST$_{n}$ cannot exceed the maximum power limit, denoted as $\bar{P}_{t}$. Under the constraints of the above two requirements, we formulate an  optimization problem to maximize the sum throughput of the STs. 

\subsection{Numerical Results} 
 
We consider the PT as a small cell base station with 1 W transmit power. The bandwidth and the frequency of the primary channel are 100 kHz and 900 MHz, respectively.  The antenna gains of the PT, STs, and SAP are all set to 6 dBi. For ambient backscatter communications, we consider a 100 kbps transmission rate. The circuit power consumption $E_{C}$, minimal demand threshold $R_{t}$ and maximum transmit power $\bar{P}_{t}$ of each ST are -25 dBm, 10 kbps and 1 W, respectively.  The energy conversion efficiency $\mu$ and data transmission efficiency $\eta$ are set at $0.15$ and $0.6$, respectively. The power of additive white Gaussian noise is $10^{-10}$ Watt. In addition, we use the CVX toolbox in
MATLAB to solve the formulated sum throughput maximization problem.

\begin{figure} 
\centering
 \subfigure [Sum throughput versus $\rho$ and $\beta_{1}$ ($\alpha_{1}=0.5$)] {\label{throughput1}
 \centering
 \includegraphics[width=0.45 \textwidth]{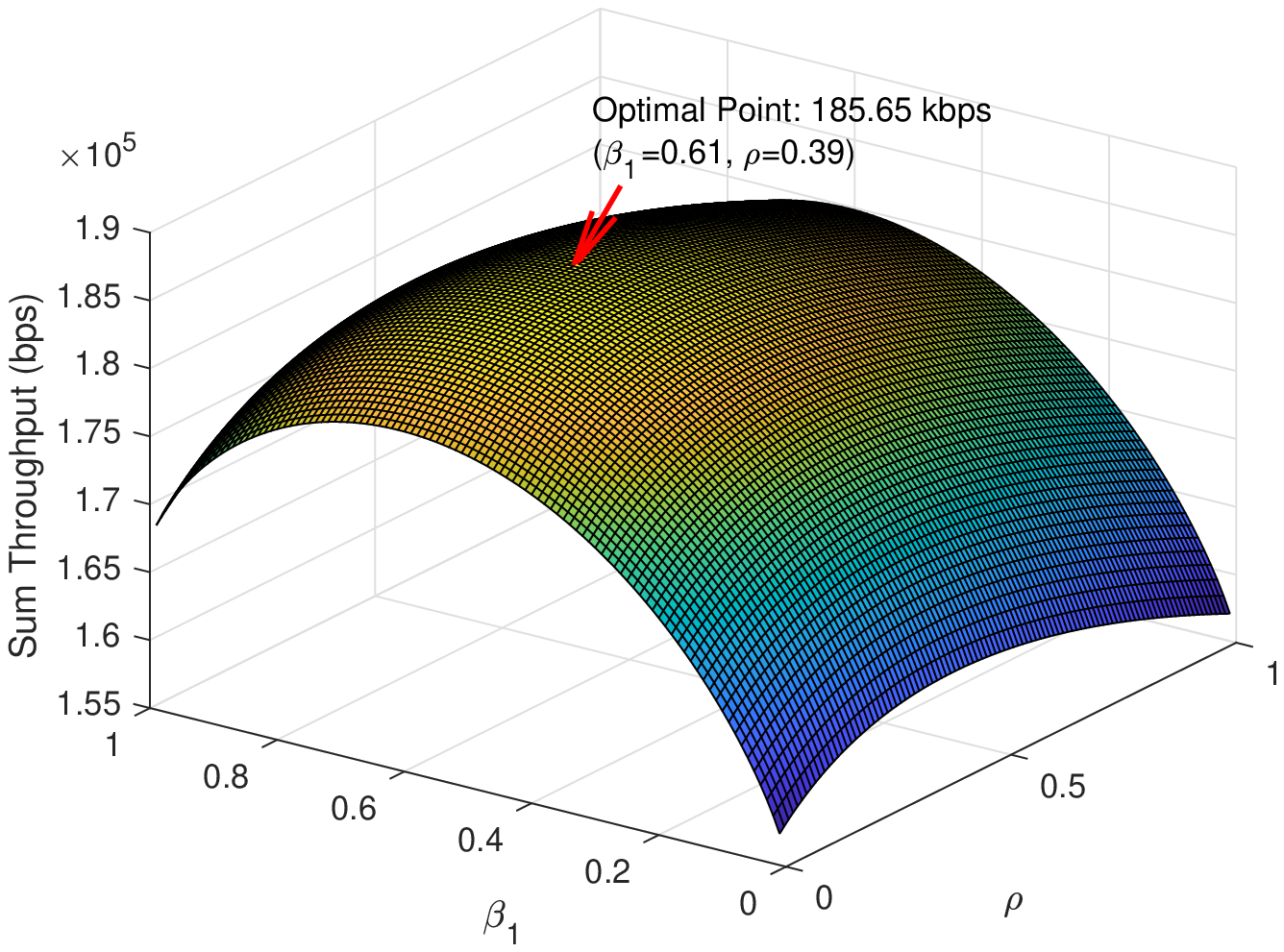}}  
 \centering
 \subfigure [Sum throughput versus $\alpha_{1}$ and $\beta_{1}$ ($\rho=0.5$)] {
 \label{throughput2}
 \centering
\includegraphics[width=0.45 \textwidth]{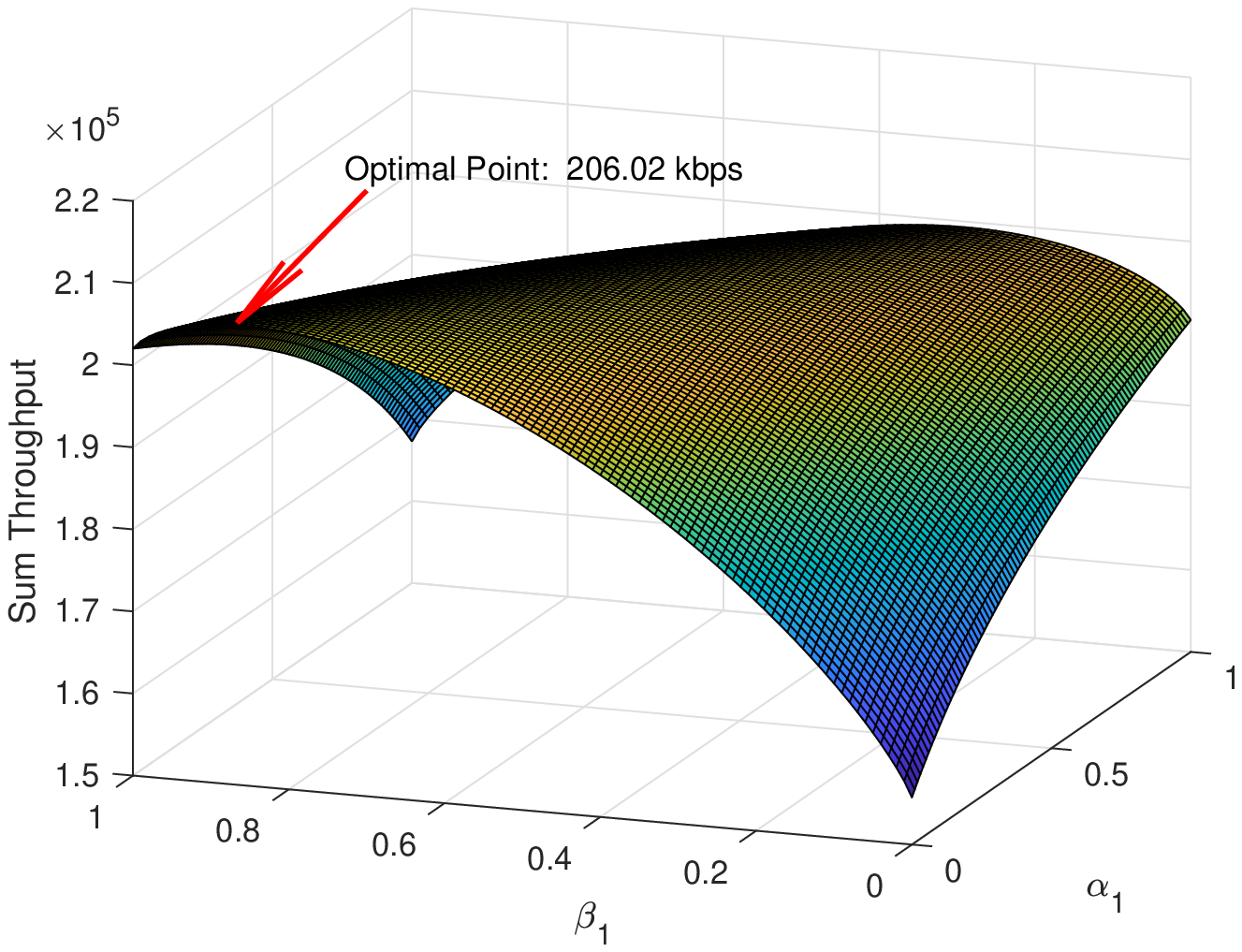}}\\
\caption{Sum throughput of two secondary transmitters with varying  $\rho$, $\alpha_1$, and $\beta_{1}$.}
\centering
\label{fig:throughput}
\end{figure}

For ease of presentation, we first consider $N=2$ STs. 
We first show how the sum throughput changes with varying $\beta_{1}$ and $\rho$ when the channel busy duration $\tau$ is 0.7. 
Both STs are placed at 11 m away from the PT. The distance between ST$_1$ and the SAP is 2 m, and that between ST$_2$ and the SAP is 2.5 m. Figure~\ref{fig:throughput}(\subref{throughput1}) shows the sum throughput versus $\rho$ and $\beta_{1}$, when $\alpha_{1}=\alpha_{2}=0.5$, i.e., ST$_1$ and ST$_2$ have equal backscatter duration. We observe that when the energy harvesting time fraction $\rho$  varies from 0 to 1, the sum throughput first increases and then decreases after reaching a maximum value. 
The maximum sum throughput is achieved when $\beta_{1}$ is greater than 0.5. The reason is that ST$_1$ is located nearer to the SAP than ST$_{2}$, and consequently ST$_1$ has better active transmission rate. Therefore, to achieve better sum throughput, more active transmission time should be allocated to ST$_1$ once the minimal throughput requirement of ST$_2$ is met. 
Then, in Figure~\ref{fig:throughput}(\subref{throughput2}), we  investigate how $\beta_{1}$ should be taken with the variation of $\alpha_{1}$ to maximize the sum throughput  
 when $\rho$ is fixed at $\rho=0.5$. We observe that when $\alpha_{1}$ takes a small value (e.g., close to 0), 
the maximal sum throughput is achieved by choosing a large value of $\beta_{1}$. On the other hand, when $\alpha_{1}$ takes a large value (e.g., close to 1), the maximal sum throughput can be obtained by assigning a small value of $\beta_{1}$. 

Fig.~\ref{throughput1} and \ref{throughput2} show the maximal sum throughput when either $\alpha_1$ or $\rho$ is fixed. This implies that the proposed multiple access scheme, by solving our formulated optimization problem, can achieve the design goal of maximizing the sum throughput over all possible combinations of ${\bm{\alpha}}$, ${\bm{ \beta}}$, and $\rho$.


\begin{figure}
\centering
\includegraphics[width=0.4\textwidth]{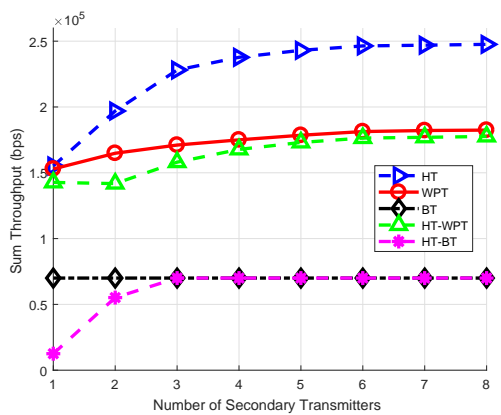} 
\caption{Sum throughput versus the number of secondary transmitters.} \label{fig:Comparison} 
\end{figure}  

In Figure~\ref{fig:Comparison}, we compare the sum throughput when the STs are (i) the proposed hybrid transmitters with the optimal solution to the formulated problem, labeled as ``HT", (ii) wireless powered transmitters only, which are equivalent to HT with $\rho=1$, labeled as ``WPT", and (iii) backscatter transmitters only, which are equivalent to HT with $\rho=0$ and $\sum_{n \in \mathcal{N}} \beta_{n}=0$, labeled as ``BT". The wireless powered transmitters harvest energy when the PT is transmitting, and sequentially perform active RF transmission when the PT is idle. The backscatter transmitters sequentially perform backscattering during the PT's transmission and remain idle when the PT is silent. For demonstration purpose, we also plot the sum active transmission throughput and sum backscattering throughput of the hybrid transmitters, labeled as ``HT-WPT" and ``HT-BT", respectively. 
We consider that the SAP is located 15 m away from the PT. 
Let $N=1,2,...,8$. The $N$ STs are aligned on the line segment between the SAP and the PT with intervals of every 1 m starting from 2~m away from the SAP towards the PT. 
As shown in Figure~\ref{fig:Comparison}, the hybrid transmitter outperforms the others by better exploiting the signal resources. In particular, the sum throughput of backscatter transmitters remains constant due to passive transmission only when the PT is transmitting. By contrast, the sum throughput of the wireless powered transmitters gradually increases as the number of STs increases because more energy is harvested by more STs for active transmission. 
Indeed, wireless powered transmitters achieve more active transmission throughput than HT-WPT. However, by exploiting ambient backscattering, the sum throughput obtained by hybrid transmitters is greater than that of the wireless powered transmitters.
Interestingly, under our parameter setting, the sum backscattering throughput of the STs remains unchanged (saturated) when there exist three or more STs. 
This is because in cases of three or more STs, an ST is scheduled to harvest energy only when another is backscattering (i.e., $\rho=0$). 
With the increase of the number of STs, the sum throughput of hybrid transmitters and that of wireless power transmitters become steady because of saturation in resource utilization.
  
\section{Future Directions and Open Issues} 

 
In this section, we discuss some future research directions for the proposed ambient backscatter assisted wireless powered communication network.  
 

\subsubsection{Integration of Ambient Backscatter with RF-powered Relay}

Similar to the idea of a hybrid transmitter, incorporating ambient backscatter capability in an RF-powered relay network would also be an interesting research direction. 
The operation of this relay network would additionally include the scheduling and resource allocation for the relay links. 

\subsubsection{Full-Duplex Wireless Powered Communications} 
 
With full-duplex operation, during the energy harvesting process, our hybrid transmitter can simultaneously perform energy harvesting and active wireless transmission~\cite{Y.2015Zeng}.
With this implementation, the optimal control of the hybrid transmitter needs to be obtained by taking into account the extra benefit of full-duplex operation.
 
\subsubsection{Multiple Access Scheme for the Hybrid Transmitters with Multiple Channels in Cognitive Radio Networks}

The performance of the hybrid transmitters can be potentially improved in a network with multiple PTs and multiple licensed channels, because there are richer signal resources for energy harvesting and more possible spectrum holes to allow concurrent active data transmissions. 
This dynamic environment complicates the coordination among the hybrid transmitters as it introduces an additional tradeoff among energy harvesting, backscattering, active transmission and channel selection. 

\section{Conclusions} 

Ambient backscatter communication technology, emerging as an integration of wireless backscatter technology and RF energy harvesting based on existing radio resources, enables sustainable and ubiquitous connections among small devices, paving the way for the development of IoT. In this article, we have proposed and evaluated a hybrid transmitter that combines ambient backscatter and wireless powered communication capabilities. Through numerical simulation and comparison, we have demonstrated the superiority of the hybrid transmitter. With self-sustainability and licensed channel-free operation, our design is suitable for low-power data communication applications.
The tradeoff among energy harvesting, active transmission and ambient backscatter communication has been studied. Future research issues that can be further addressed have been discussed.
 

\end{document}